\documentclass[aps,a4paper,superscriptaddress,preprintnumbers,showpacs,amsmath,amssymb]{revtex4}
\usepackage{graphicx}
\usepackage{dcolumn}
\usepackage{color}
\usepackage{latexsym,amsfonts}

\usepackage{textcomp}

\usepackage{bm}

\usepackage{ulem}

\begin{document}

\title{Non--Relativistic Approximation of Dirac Equation\\ for Slow
  Fermions  Coupled to the Chameleon and Torsion Fields\\ in the
  Gravitational Field of the Earth}

\author{A. N. Ivanov}\email{ivanov@kph.tuwien.ac.at}
\affiliation{Atominstitut, Technische Universit\"at Wien, Stadionallee
  2, A-1020 Wien, Austria}
\author{M. Wellenzohn}\email{max.wellenzohn@gmail.com}
\affiliation{Atominstitut, Technische Universit\"at Wien, Stadionallee
  2, A-1020 Wien, Austria}
\affiliation{FH Campus Wien, University of Applied Sciences, 
Favoritenstra\ss e 226, 1100 Wien, Austria}

\date{\today}

\begin{abstract}
We analyse a non--relativistic approximation of the Dirac equation for
slow fermions, coupled to the chameleon field and torsion in the
spacetime with the Schwarzschild metric, taken in the weak
gravitational field of the Earth approximation. We follow the analysis
of the Dirac equation in the curved spacetime with torsion, proposed
by Kostelecky (Phys. Rev. D {\bf 69}, 105009 (2004)), and apply the
Foldy--Wouthuysen transformations. We derive the effective low--energy
gravitational potentials for slow fermions, coupled to the
gravitational field of the Earth, the chameleon field and to torsion
with minimal and non--minimal couplings.
\end{abstract}

\maketitle

\section{Introduction}
\label{sec:introduction}

Recently the non--relativistic approximation of the Dirac equation for
slow fermions, moving in the gravitational field of the Earth,
described by the Schwarzschild metric in the weak gravitational field
approximation, has been analysed in Ref.\cite{Ivanov2014}. As has been
shown in \cite{Ivanov2014}, slow fermions couple to the external
gravitational field through the effective low--energy gravitational
potential
\begin{eqnarray}\label{eq:1}
  \Phi_{\rm eff}(\vec{r}, \vec{\nabla}, \vec{\sigma}\,) =
  m\,U(\vec{r}\,) - \frac{1 + 2\gamma}{2m}\,\Big(U(\vec{r}\,)\,\Delta
  + \vec{\nabla} U(\vec{r}\,)\cdot \vec{\nabla} + \frac{1}{4}\,\Delta
  U(\vec{r}\,)\Big) - i\frac{1 +
    2\gamma}{4m}\,\vec{\sigma}\cdot(\vec{\nabla}U(\vec{r}\,)\times
  \vec{\nabla}\,),
\end{eqnarray}
where $\vec{r}$ is the radius vector of a slow fermion, $\vec{\nabla}
= \partial/\partial \vec{r}$ is the gradient and $U(\vec{r}\,)$ is the
gravitational potential taken in the form
\begin{eqnarray}\label{eq:2}
  U(\vec{r}\,) = \vec{g}\cdot \vec{r} + \frac{\beta}{M_{\rm
      Pl}}\,\phi(\vec{r}\,).
\end{eqnarray}
The first term is the Newtonian gravitational potential of the Earth
with the gravitational acceleration $\vec{g}$, whereas the second one
stands for the contribution of the chameleon field $\phi(\vec{r}\,)$
and describes a deviation from the Newtonian gravity
\cite{Brax2011,Ivanov2013}. It is determined by the reduced Planck
mass $M_{\rm Pl} = 1/\sqrt{8\pi G_N} = 2.435\times 10^{27}\,{\rm eV}$,
where $G_N $ is the gravitational constant \cite{PDG2014}, and the
chameleon--matter coupling constant $\beta < 5.8\times 10^8$
\cite{Jenke2014}. The last term in Eq.(\ref{eq:1}) has been
interpreted as the potential of the torsion--fermion interaction with
the torsion field $\vec{\cal T} = (\beta/M_{\rm
  Pl})\,\vec{\nabla}\phi(\vec{r}\,)$, caused by the chameleon
field. For the confirmation of the relation of the last term in
Eq.(\ref{eq:1}) to the torsion field there have been used the results,
obtained by Kostelecky {\it et al.}
\cite{Kostelecky2008,Kostelecky2011}. An extension of the results,
obtained in \cite{Ivanov2014}, as a version of the Einstein--Cartan
gravitational theory with torsion, defined by the gradient of the
chameleon field, has been proposed in \cite{Ivanov2015}.

In this paper we give a derivation of the effective low--energy
gravitational potential of slow fermions, coupled to the chameleon,
torsion and weak gravitational field in the spacetime
  defined by the Schwarzschild metric. In the Einstein frame it takes
the form
\begin{eqnarray}\label{eq:3}
  ds^2 = g_{\mu\nu}(x)dx^{\mu}dx^{\nu} = (1 + 2 U_{\rm E})\,dt^2 - (1
  - 2 U_{\rm E})\,d\vec{r}^{\;2},
\end{eqnarray}
where $U_{\rm E}$ is a gravitational potential of the Earth. The
torsion field ${{\cal T}^{\alpha}}_{\mu\nu} = - {{\cal
    T}^{\alpha}}_{\nu\mu}$ is related to the affine connection as follows
\cite{Hojman1978} (see also \cite{Ivanov2015})
\begin{eqnarray}\label{eq:4}
{\Gamma^{\alpha}}_{\mu\nu} = \{{^{\alpha}}_{\mu\nu}\} - \frac{1}{2}\,
({{\cal T}^{\alpha}}_{\mu\nu} - {{{\cal T}_{\mu}}^{\alpha}}_{\nu} -
{{{\cal T}_{\nu}}^{\alpha}}_{\mu}) = \{{^{\alpha}}_{\mu\nu}\} +
 g^{\alpha\sigma}{\cal K}_{\sigma\mu\nu},
\end{eqnarray}
where $\{{^{\alpha}}_{\mu\nu}\}$ are the Christoffel symbols
\cite{Fliessbach2006,Rebhan2012}
\begin{eqnarray}\label{eq:5}
\{{^{\alpha}}_{\mu\nu}\} =
\frac{1}{2}g^{\alpha\lambda}\Big(\frac{\partial
  g_{\lambda\mu}}{\partial x^{\nu}} + \frac{\partial
  g_{\lambda\nu}}{\partial x^{\mu}} - \frac{\partial
  g_{\mu\nu}}{\partial x^{\lambda}}\Big)
\end{eqnarray}
and ${\cal K}_{\sigma\mu\nu} = - (1/2)({\cal T}_{\sigma\mu\nu} - {\cal
  T}_{\mu\sigma\nu} - {\cal T}_{\nu\sigma\mu})$ is the contorsion
tensor. The metric tensors $g_{\alpha\beta}$ and $g^{\alpha\beta}$
have the following components: $g_{00} = 1 + 2U_{\rm E}$, $g_{0j} = 0$
and $g_{ij} = - (1 - 2 U_{\rm E})\,\delta_{ij}$ and $g^{00} = 1 -
2U_{\rm E}$, $g^{0j} = 0$ and $g^{ij} = - (1 + 2 U_{\rm
  E})\,\delta_{ij}$ \cite{Ivanov2014}, respectively, taken in the
linear approximation with respect to the $U_{\rm E}$--expansion.

An analogous analysis of the Dirac equation for slow fermions, moving
in spacetime with the Schwarzschild metric in a weak gravitational
field approximation and torsion but without the contribution of the
chameleon field, has been carried out by L\"ammerzahl \cite{
  Laemmerzahl1997}. Unlike L\"ammerzahl \cite{ Laemmerzahl1997} we
take into account a dependence of torsion on a
  spacetime point $x$.  We compare our results also with the results,
obtained by Obukhov, Silenko, and Teryaev \cite{Obukhov2014}, who have
discussed a quantum dynamics of a Dirac fermion in the
Poincar$\acute{\rm e}$ gauge gravitational field, incorporating a
torsion field. In addition we have derived the effective low--energy
gravitational potential for slow fermions, coupled non--minimally to
the torsion field. Throughout the paper we follow the analysis of the
Dirac equation in the curved spacetime with torsion, proposed by
Kostelecky \cite{Kostelecky2004}.

The paper is organized as follows. In section \ref{sec:curvespace} we
give a general expression for the Dirac equation in terms of the
vierbein fields, the spin connection and torsion. In section
\ref{sec:hamilton} we calculate the Dirac Hamilton operator for slow
fermions to linear order of interacting gravitational, chameleon and
torsion fields. In section \ref{sec:foldy} we make the
Foldy--Wouthuysen transformations and derive the effective low--energy
gravitational potential for slow fermions, coupled to the chameleon
field and torsion with minimal couplings in the spacetime with the
Schwarzschild metric, taken in the weak gravitational field of the
Earth approximation. In section \ref{sec:torsion} we derive the
effective low--energy gravitational potential for slow fermions,
coupled non--minimally to torsion. In section \ref{sec:conclusion} we
discuss the obtained results. In the Appendix we give a detailed
derivation of the Dirac equation and the Dirac Hamilton operator for
Dirac fermions in the curved spacetime with torsion.

\section{Dirac equation for slow fermions in curved spacetime with 
a weak gravitational field, the chameleon field and torsion}
\label{sec:curvespace}

In the gravitational theory with the chameleon field slow fermions
couple to the chameleon field through the metric $\tilde{g}_{\mu\nu}$
in the Jordan frame related to the metric $g_{\mu\nu}$ in the Einstein
frame by $\tilde{g}_{\mu\nu} = f^2\,g_{\mu\nu}$, where $f =
e^{\,\beta\phi/M_{\rm Pl}}$ is the conformal factor
\cite{Chameleon1,Chameleon2} (see also \cite{Ivanov2015}).

The analysis of the Dirac equation for slow fermions coupled to the
chameleon field in the spacetime with torsion and the metric
$d\tilde{s}^2 = \tilde{g}_{\mu\nu}(x)dx^{\mu}dx^{\nu}$ (the
Jordan--frame metric) we start with the action \cite{Kostelecky2004}
(see also the Appendix)
\begin{eqnarray}\label{eq:6}
\hspace{-0.3in}{\rm S}_{\psi} = \int d^4x\,\sqrt{-
  \tilde{g}}\,\Big(i\,\frac{1}{2}\,\bar{\psi}(x)\tilde{\gamma}^{\mu}(x)\!
\stackrel{\leftrightarrow}{D}_{\mu}\!\psi(x) - m\bar{\psi}(x)\psi(x)\Big),
\end{eqnarray}
where $m$ is the fermion mass, $\tilde{g} = {\rm
  det}\{\tilde{g}_{\mu\nu}\}$ is the determinant of the metric tensor
and $\tilde{\gamma}^{\mu}(x)$ are the Dirac matrices in the Jordan
frame satisfying the anticommutation relation
\begin{eqnarray}\label{eq:7}
\hspace{-0.3in}\tilde{\gamma}^{\mu}(x)\tilde{\gamma}^{\nu}(x) +
\tilde{\gamma}^{\nu}(x)\tilde{\gamma}^{\mu}(x) = 2 \tilde{g}^{\mu\nu}(x)
\end{eqnarray}
and $D_{\mu}$ is a covariant derivative without gauge fields. For an
exact definition of the Dirac matrices $\tilde{\gamma}^{\mu}(x)$ and
the covariant derivative $D_{\mu}$ we follow
\cite{Kostelecky2004,Fischbach1981} and use a set of vierbein fields
$\tilde{e}^{\hat{\alpha}}_{\mu}(x)$ at each spacetime point $x$
defined by
\begin{eqnarray}\label{eq:8}
\hspace{-0.3in}dx^{\hat{\alpha}} = \tilde{e}^{\hat{\alpha}}_{\mu}(x) dx^{\mu}.
\end{eqnarray}
The vierbein fields relate in an arbitrary (world) coordinate system a
spacetime point $x$, which is characterized by the index $\mu =
0,1,2,3$, to a locally Minkowskian coordinate system erected at a
spacetime point $x$, which is characterized by the index $\hat{\alpha}
= 0,1,2,3$. The vierbein fields $\tilde{e}^{\hat{\alpha}}_{\mu}(x)$ are related
to the metric tensor $\tilde{g}_{\mu\nu}(x)$ by
\begin{eqnarray}\label{eq:9}
\hspace{-0.3in}d\tilde{s}^2 = \eta_{\hat{\alpha}\hat{\beta}}
\,dx^{\hat{\alpha}} dx^{\hat{\beta}} =
\eta_{\hat{\alpha}\hat{\beta}}\,[\tilde{e}^{\hat{\alpha}}_{\mu}(x)dx^{\mu}]
    [\tilde{e}^{\hat{\beta}}_{\nu}(x)dx^{\nu}] =
    [\eta_{\hat{\alpha}\hat{\beta}}\,\tilde{e}^{\hat{\alpha}}_{\mu}(x)
      \tilde{e}^{\hat{\beta}}_{\nu}(x)] dx^{\mu}dx^{\nu} =
    \tilde{g}_{\mu\nu}(x)dx^{\mu}dx^{\nu},
\end{eqnarray}
where $\eta_{\hat{\alpha}\hat{\beta}}$ is the metric tensor in the
Minkowski spacetime with the signature $\eta_{\hat{\alpha}\hat{\beta}}
= (1,-1,-1,-1)$. This gives
\begin{eqnarray}\label{eq:10}
\hspace{-0.3in} \tilde{g}_{\mu\nu}(x) =
\eta_{\hat{\alpha}\hat{\beta}}\,\tilde{e}^{\hat{\alpha}}_{\mu}(x)
\tilde{e}^{\hat{\beta}}_{\nu}(x).
\end{eqnarray}
Thus, the vierbein fields can be viewed as the square root of the metric
tensor $\tilde{g}_{\mu\nu}(x)$ in the sense of a matrix equation
\cite{Fischbach1981}. Inverting the relation Eq.(\ref{eq:8}) we
obtain
\begin{eqnarray}\label{eq:11}
\hspace{-0.3in} \eta_{\hat{\alpha}\hat{\beta}}= \tilde{g}_{\mu\nu}(x)
\tilde{e}^{\mu}_{\hat{\alpha}}(x) \tilde{e}^{\nu}_{\hat{\beta}}(x).
\end{eqnarray}
There are also the following relations
\begin{eqnarray}\label{eq:12}
\tilde{e}^{\mu}_{\hat{\alpha}}(x) \tilde{e}^{\hat{\beta}}_{\mu}(x) &=&
\delta^{\hat{\beta}}_{\hat{\alpha}},\nonumber\\ \tilde{e}^{\mu}_{\hat{\alpha}}(x)
\tilde{e}^{\hat{\alpha}}_{\nu}(x) &=&
\delta^{\mu}_{\nu},\nonumber\\ \tilde{e}^{\mu}_{\hat{\alpha}}(x)
\tilde{e}_{\hat{\beta}\mu}(x) &=&
\eta_{\hat{\alpha}\hat{\beta}},\nonumber\\ \tilde{e}_{\hat{\alpha}\mu}(x)
&=&
\eta_{\hat{\alpha}\hat{\beta}}\,\tilde{e}^{\hat{\beta}}_{\mu}(x),\nonumber\\ \tilde{e}^{\hat{\alpha}}_{\mu}(x)
\tilde{e}_{\hat{\alpha}\nu}(x) &=& \tilde{g}_{\mu\nu}(x),
\end{eqnarray}
which are useful for the derivation of the Dirac equation and
calculation of the Dirac Hamilton operator of fermions with mass $m$
(see the Appendix).  In terms of the vierbein fields the Dirac
matrices $\tilde{\gamma}^{\mu}(x)$ are defined by
\begin{eqnarray}\label{eq:13}
\hspace{-0.3in} \tilde{\gamma}^{\mu}(x) =
\tilde{e}^{\mu}_{\hat{\alpha}}(x) \gamma^{\hat{\alpha}},
\end{eqnarray}
where $\gamma^{\hat{\alpha}}$ are the Dirac matrices in the Minkowski
spacetime \cite{Itzykson1980}. A covariant derivative $D_{\mu}$ we
define as \cite{Kostelecky2004}
\begin{eqnarray}\label{eq:14}
\hspace{-0.3in} D_{\mu} = \partial_{\mu} - \tilde{\Gamma}_{\mu}(x),
\end{eqnarray}
where $\tilde{\Gamma}_{\mu}(x)$ is the spin affine connection. In
terms of the spin connection
$\tilde{\omega}_{\mu\hat{\alpha}\hat{\beta}}(x)$ it is given by
\cite{Kostelecky2004}
\begin{eqnarray}\label{eq:15}
\hspace{-0.3in} \tilde{\Gamma}_{\mu}(x) =
\frac{i}{4}\,\tilde{\omega}_{\mu\hat{\alpha}\hat{\beta}}
\sigma^{\hat{\alpha}\hat{\beta}},
\end{eqnarray}
where $\sigma^{\hat{\alpha}\hat{\beta}} = (i/2)
(\gamma^{\hat{\alpha}}\gamma^{\hat{\beta}} -
\gamma^{\hat{\beta}}\gamma^{\hat{\alpha}})$ are the Dirac matrices in
the Minkowski spacetime \cite{Itzykson1980}. The derivation of the
Dirac equation in the curved spacetime with the metric tensor
$\tilde{g}_{\mu\nu}(x)$ and torsion we have carried out in the
Appendix. The result is
\begin{eqnarray}\label{eq:16}
\Big( i\,\tilde{e}^{\mu}_{\hat{\lambda}}(x) \gamma^{\hat{\lambda}}
D_{\mu} - \frac{1}{2}\,i\,{\tilde{\cal
    T}^{\alpha}\,\!\!}_{\alpha\mu}(x)
\tilde{e}^{\mu}_{\hat{\lambda}}(x) \gamma^{\hat{\lambda}} -
\frac{1}{2}\,i\, \tilde{\omega}_{\mu\hat{\alpha}\hat{\beta}}(x)
\tilde{e}^{\mu}_{\hat{\lambda}}(x)
\Big(\eta^{\hat{\lambda}\hat{\beta}}\gamma^{\hat{\alpha}} +
\frac{1}{4}\,i\, [\sigma^{\hat{\alpha}\hat{\beta}},
  \gamma^{\hat{\lambda}}]\Big) - m\Big)\,\psi(x) = 0,
\end{eqnarray}
where $[\sigma^{\hat{\alpha}\hat{\beta}}, \gamma^{\hat{\lambda}}] =
\sigma^{\hat{\alpha}\hat{\beta}} \gamma^{\hat{\lambda}} -
\gamma^{\hat{\lambda}} \sigma^{\hat{\alpha}\hat{\beta}}$. The Dirac
equation Eq.(\ref{eq:16}) agrees well with Eq.(\ref{eq:18}) of
Ref.\cite{Kostelecky2004}.  The spin connection
$\tilde{\omega}_{\mu\hat{\alpha}\hat{\beta}}(x)$ is related to the
vierbein fields and the affine connection as follows
\cite{Kostelecky2004}
\begin{eqnarray}\label{eq:17}
\tilde{\omega}_{\mu\hat{\alpha}\hat{\beta}}(x) = -
\eta_{\hat{\alpha}\hat{\varphi}}\Big(\partial_{\mu}\tilde{e}^{\hat{\varphi}}_{\nu}(x)
- {\tilde{\Gamma}^{\alpha}\,}_{\mu\nu}(x)
\tilde{e}^{\hat{\varphi}}_{\alpha}(x)\Big)\tilde{e}^{\nu}_{\hat{\beta}}(x).
\end{eqnarray}
The vierbein fields $\tilde{e}^{\hat{\alpha}}_{\mu}(x)$ in the Jordan
frame are related to the vierbein fields $e^{\hat{\alpha}}_{\mu}(x)$
in the Einstein frame by \cite{Ivanov2015}
\begin{eqnarray}\label{eq:18}
\tilde{e}^{\hat{\alpha}}_{\mu}(x) = f\,e^{\hat{\alpha}}_{\mu}(x)\;,\;
    \tilde{e}^{\mu}_{\hat{\alpha}}(x) = f^{-1}\,e^{\mu}_{\hat{\alpha}}(x).
\end{eqnarray}
For the Einstein--frame metric Eq.(\ref{eq:3}) the vierbein fields are
equal to
\begin{eqnarray}\label{eq:19}
\hspace{-0.3in} e^{\hat{\alpha}}_0(x) &=& (1 +
U_{\rm E})\,\delta^{\hat{\alpha}}_0\;,\; e^0_{\hat{\alpha}}(x) = (1 -
U_{\rm E})\,\delta^0_{\hat{\alpha}},\nonumber\\
e^{\hat{\alpha}}_j(x) &=& (1 -
U_{\rm E})\,\delta^{\hat{\alpha}}_j\;,\; e^j_{\hat{\alpha}}(x) = (1 +
U_{\rm E})\,\delta^j_{\hat{\alpha}},
\end{eqnarray}
where we have kept only the linear terms in the $U_{\rm
  E}$--expansion. Expanding the conformal factor $f =
e^{\,\beta\phi/M_{\rm Pl}}$ in powers of $\beta\phi/M_{\rm Pl}$ we
define the vierbein fields $\tilde{e}^{\hat{\alpha}}_{\mu}(x)$ in the
Jordan frame as
\begin{eqnarray}\label{eq:20}
\hspace{-0.3in} \tilde{e}^{\hat{\alpha}}_0(x) &=& (1 +
U_+)\,\delta^{\hat{\alpha}}_0\;,\; \tilde{e}^0_{\hat{\alpha}}(x) = (1 -
U_+)\,\delta^0_{\hat{\alpha}},\nonumber\\
\tilde{e}^{\hat{\alpha}}_j(x) &=& (1 -
U_-)\,\delta^{\hat{\alpha}}_j\;,\; \tilde{e}^j_{\hat{\alpha}}(x) = (1 +
U_-)\,\delta^j_{\hat{\alpha}}.
\end{eqnarray}
The potentials $U_{\pm}$ are determined by
\begin{eqnarray}\label{eq:21}
U_{\pm} = U_{\rm E} \pm  \frac{\beta}{M_{\rm Pl}}\,\phi,
\end{eqnarray}
where we have kept only the linear contributions of the chameleon
field and the gravitational field of the Earth. In
such an approximation the spin affine connection
$\tilde{\Gamma}_{\mu}(x)$ is defined by the affine connection as
follows
\begin{eqnarray}\label{eq:22}
{\tilde{\Gamma}^{\alpha}\,}_{\mu\nu} =
\frac{1}{2}\,\tilde{g}^{\alpha\lambda}\Big(\frac{\partial
  \tilde{g}_{\lambda\mu}}{\partial x^{\nu}} + \frac{\partial
  \tilde{g}_{\lambda\nu}}{\partial x^{\mu}} - \frac{\partial
  \tilde{g}_{\mu\nu}}{\partial x^{\lambda}}\Big) - \frac{1}{2}\,
({{\cal T}^{\alpha}}_{\mu\nu} - {{{\cal T}_{\mu}}^{\alpha}}_{\nu} -
     {{{\cal T}_{\nu}}^{\alpha}}_{\mu}),
\end{eqnarray}
where $\tilde{g}_{00} = 1 + 2 U_+$, $\tilde{g}_{0j} = \tilde{g}_{j0} =
0$, $\tilde{g}_{ij} = (1 - 2 U_-)\,\eta_{ij}$ and $\eta_{ij} = -
\delta_{ij}$ determine the Schwarzschild metric in a weak
gravitational field approximation modified by the contribution of the
chameleon field. In the linear approximation for the definition of the
contribution of the torsion field we have set the conformal factor $f
= 1$. Thus, Eq.(\ref{eq:16}) with the vierbein fields, given by
Eq.(\ref{eq:20}), and the spin affine connection, determined in terms
of the affine connection Eq.(\ref{eq:22}), is the Dirac equation for
fermions, coupled to the torsion field ${{\cal T}^{\alpha}}_{\mu\nu}$
in the spacetime with the Schwarzschild metric: $\tilde{g}_{00} = 1 +
2 U_+$, $\tilde{g}_{0j} = \tilde{g}_{j0} = 0$ and $\tilde{g}_{ij} = -
(1 - 2 U_-)\,\delta_{ij}$, modified by the contribution of the
chameleon field. The Dirac equation Eq.(\ref{eq:16}) realizes also a
minimal coupling for the torsion--fermion (matter) interactions.

\section{Dirac Hamilton operator for slow fermions 
in the gravitational field of the Earth with chameleon and torsion
fields}
\label{sec:hamilton}

In the Appendix we have derived the general expression for the Dirac
Hamilton operator for fermions in the curved space time with the
chameleon field and torsion. For the derivation of an effective
gravitational potential for slow fermions we have approximated such a
Hamilton operator keeping only the linear order contributions of
interacting fields. The Dirac equation for slow fermions in the
standard form is
\begin{eqnarray}\label{eq:23}
i\,\frac{\partial \psi}{\partial t}= {\rm H}\,\psi,
\end{eqnarray}
where ${\rm H} = {\rm H}_0 + \delta {\rm H}$ and ${\rm H}_0 =
\gamma^{\hat{0}}m - i\,\gamma^{\hat{0}}\,\hat{\!\vec{\gamma}}\cdot
\vec{\nabla}$ is the Hamilton operator of free fermions, whereas
$\delta {\rm H}$ defines the interactions of fermions with
gravitational, chameleon and torsion fields
\begin{eqnarray}\label{eq:24}
\hspace{-0.3in}\delta {\rm H} &=& (\tilde{e}^{\hat{0}}_0(x) - 1)
\gamma^{\hat{0}} m - i\,(\tilde{e}^{\hat{0}}_0(x) -
1)\gamma^{\hat{0}}\gamma^{\hat{j}}\delta^j_{\hat{j}}\frac{\partial
}{\partial x^j} - i\tilde{e}^{\hat{0}}_0(x)(\tilde{e}^j_{\hat{j}}(x) -
\delta^j_{\hat{j}})\gamma^{\hat{0}}\gamma^{\hat{j}}\frac{\partial
}{\partial x^j} -
\frac{1}{2}\,i\,\eta^{\mu\nu}\Big(\partial_{\mu}\tilde{e}^{\hat{\lambda}}_{\nu}(x)
-
\{{^{\hat{\lambda}}}_{\mu\nu}\}\Big)\gamma^{\hat{0}}\gamma_{\hat{\lambda}}\nonumber\\
\hspace{-0.3in}&-& 
\frac{1}{8}\,\epsilon^{\hat{\alpha}\hat{\beta}\mu\nu}{\cal
  T}_{\hat{\beta}\mu\nu}\gamma^{\hat{0}}\gamma_{\hat{\alpha}}\gamma^5
\end{eqnarray}
to linear order of interacting fields, where
$\epsilon^{\hat{\alpha}\hat{\beta}\mu\nu}$ is the Levy--Civita tensor
such as $\epsilon^{0123} = +1$ and $\epsilon^{0j\ell m} =
\epsilon^{j\ell m}$ and $\gamma^5 = i \gamma^{\hat{0}}
\gamma^{\hat{1}} \gamma^{\hat{2}} \gamma^{\hat{3}}$ is the Dirac
matrix \cite{Itzykson1980}. We would like to note that below because
of the proportionality of the vierbein fields to Kronecker tensors
Eq.(\ref{eq:20}) we do not distinguish the indices in the Minkowski
spacetime $\hat{\alpha}$ and the indices in the curved spacetime
$\alpha$. This is also confirmed by the use of the weak gravitational,
torsion and chameleon field approximation, where the gravitational,
torsion and chameleon fields appear in the form of interactions with
Dirac fermions in the Minkowski spacetime.  Using the vierbein fields
Eq.(\ref{eq:20}) and the relation
\begin{eqnarray}\label{eq:25}
\hspace{-0.3in} -
\frac{1}{2}\,i\,\eta^{\mu\nu}\Big(\partial_{\mu}\tilde{e}^{\hat{\lambda}}_{\nu}
- \{{^{\hat{\lambda}}}_{\mu\nu}\}\Big)
  \gamma^{\hat{0}}\gamma_{\hat{\lambda}} =
  i\,\frac{3}{2}\,\frac{\partial U_-}{\partial t} -
  \frac{1}{2}\,i\,\gamma^0\vec{\gamma}\cdot \vec{\nabla}U_+ +
  i\,\gamma^0\vec{\gamma}\cdot \vec{\nabla}U_-
\end{eqnarray}
we transcribe the Hamilton operator $\delta {\rm H}$ into the form
\begin{eqnarray}\label{eq:26}
\hspace{-0.3in}\delta {\rm H} = i\,\frac{3}{2}\,\frac{\partial
  U_-}{\partial t} + \gamma^0 U_+ m - i\,(U_+ + U_-)\, \gamma^0
\vec{\gamma}\cdot \vec{\nabla} -
\frac{1}{2}\,i\,\gamma^0\vec{\gamma}\cdot \vec{\nabla}U_+ +
i\,\gamma^0\vec{\gamma}\cdot \vec{\nabla}U_- - \frac{1}{4}\,\gamma^5\,{\cal K} - \frac{1}{4}\,\vec{\Sigma}\cdot \vec{\cal B},
\end{eqnarray}
where we have denoted $\vec{\Sigma} = \gamma^0 \vec{\gamma} \gamma^5$,
which is the $4\times 4$--diagonal matrix with elements defined by the
$2\times 2$ Pauli matrices $\vec{\sigma}$ \cite{Itzykson1980}, and
\begin{eqnarray}\label{eq:27}
 {\cal K} &=& \frac{1}{2}\,\epsilon^{j\ell m}\,{\cal T}_{j\ell
   m},\nonumber\\ (\vec{\cal B}\,)^j &=& \frac{1}{2}\,\epsilon^{j
   \ell m }({\cal T}_{\ell m 0} + {\cal T}_{m 0 \ell} + {\cal T}_{0
   \ell m}),
\end{eqnarray}
which are the time and space components of the axial 4--vector ${\cal
  B}^{\alpha} = \frac{1}{2}\,\epsilon^{\alpha\beta\mu\nu}{\cal
  T}_{\beta\mu\nu} = ({\cal K}, \vec{\cal B}\,)$,
respectively. 

For the subsequent derivation of the low--energy Hamilton operator of
slow fermions we have to make the standard transformations of the wave
function of slow fermions and the Hamilton operator
\cite{Obukhov2014,Fischbach1981} (see also \cite{Ivanov2014})
\begin{eqnarray}\label{eq:28}
\psi(x)&=& (1 + \frac{3}{2}\,U_-)\,\psi'(x)\nonumber\\ {\rm H}' &=&
    {\rm H}_0 + \delta {\rm H} + \frac{3}{2}\,[{\rm H}_0, U_-] -
    i\,\frac{3}{2}\,\frac{\partial U_-}{\partial t},
\end{eqnarray}
where we have kept only the linear contributions of gravitational,
chameleon and torsion fields.  For the Hamilton operator ${\rm H}'$ we
obtain the following expression
\begin{eqnarray}\label{eq:29}
{\rm H}' = {\rm H}_0 + \delta {\rm H}',
\end{eqnarray}
where $\delta {\rm H}'$ is given by
\begin{eqnarray}\label{eq:30}
\delta {\rm H}' =\gamma^0\,m\,U_+ - i\,(U_+ +
  U_-)\,\gamma^0\vec{\gamma}\cdot \vec{\nabla} - \frac{i}{2}\,\gamma^0
  \vec{\gamma}\cdot \vec{\nabla}(U_+ + U_-) -
  \gamma^5\,\frac{1}{4}\,{\cal K} - \frac{1}{4}\,\vec{\Sigma}\cdot
  \vec{\cal B}.
\end{eqnarray}
The Dirac equation in its standard form reads
\begin{eqnarray}\label{eq:31}
\hspace{-0.15in}i\frac{\partial \psi'}{\partial t} = {\rm H}'\,\psi'.
\end{eqnarray}
The Hamilton operator Eq.(\ref{eq:30}) agrees well with those
calculated by L\"ammerzahl (see Eq.(\ref{eq:16}) of
Ref.\cite{Laemmerzahl1997}) and Obukhov, Silenko, and Teryaev (see
Eq.(2.21) of Ref.\cite{Obukhov2014}). However, these authors did not
take into account the contributions of the chameleon field.

\section{Effective Hamilton operator for slow fermions in the 
gravitational field of the Earth with chameleon and torsion fields}
\label{sec:foldy}

For the derivation of the low--energy approximation of the Dirac
equation Eq.(\ref{eq:31}) we use the Foldy--Wouthuysen (FW)
transformation \cite{Foldy1950}. The aim of the FW transformation is
to delete all {\it odd} operators, which are proportional to
$\vec{\gamma}$, $\gamma^0\vec{\gamma}$, $\gamma^5$ and
$\gamma^0\gamma^5$. As a result, the final low--energy
  Hamilton operator should be expressed in terms of the {\it even}
  operators only, which are proportional to $\gamma^0$, $\vec{\Sigma}
  = \gamma^0\,\vec{\gamma}\,\gamma^5$ and $\gamma^0 \vec{\Sigma}$,
  respectively.  For the elimination of {\it odd} operators we
perform the unitary transformation \cite{Foldy1950,Bjorken1966}
\begin{eqnarray}\label{eq:32}
{\rm H}_1 = e^{\,+ i S_1}\,{\rm H}'\,e^{\,-i S_1} - i\,e^{\,i
  S_1}\frac{\partial}{\partial t}e^{\,-i S_1} = {\rm H}' -
\frac{\partial S_1}{\partial t} + i\Big[S_1,{\rm H}' -
  \frac{1}{2}\,\frac{\partial S_1}{\partial t}\Big] +
\frac{i^2}{2}\,\Big[S_1,\Big[S_1,{\rm H}' -
    \frac{1}{3}\,\frac{\partial S_1}{\partial t}\Big]\Big] + \ldots
\end{eqnarray}
The time derivative appears because of a time dependence of the
chameleon and torsion fields. Then, following \cite{Foldy1950} we take
the operator $S_1$ in the form
\begin{eqnarray}\label{eq:33}
S_1 &=& - \frac{i}{2m}\,\gamma^0\Big(- (1 + U_-)\,i\,\gamma^0
\vec{\gamma}\cdot \vec{\nabla} -
\frac{i}{2}\,\gamma^0\vec{\gamma}\cdot \vec{\nabla}(U_+ + U_-) -
\frac{1}{4}\,\gamma^5\,{\cal K}\Big) =\nonumber\\ &=& -
\frac{1}{2m}\,(1 + U_-)\,\vec{\gamma}\cdot \vec{\nabla} -
\frac{1}{4m}\,\vec{\gamma}\cdot \vec{\nabla}(U_+ + U_-) +
\frac{i}{8m}\,\gamma^0 \gamma^5\,{\cal K}.
\end{eqnarray}
The time derivative of $S_1$ and the commutators in Eq.(\ref{eq:32})
are equal to
\begin{eqnarray}\label{eq:34}
\frac{\partial S_1}{\partial t} &=& - \frac{1}{2m}\,\frac{\partial
  U_-}{\partial t}\,\vec{\gamma}\cdot \vec{\nabla} +
\frac{i}{8m}\,\gamma^0 \gamma^5\,\frac{\partial {\cal K}}{\partial
  t},\nonumber\\ i\Big[S_1,{\rm H}' - \frac{1}{2}\,\frac{\partial
    S_1}{\partial t}\Big] &=& i(1 + U_+ +
U_-)\,\gamma^0\,\vec{\gamma}\cdot\vec{\nabla} +
\frac{i}{2}\,\gamma^0\,\vec{\gamma}\cdot \vec{\nabla}(2 U_+ + U_-) -
\frac{\gamma^0}{2m}\,\vec{\nabla}(3 U_+ + 4 U_-)\cdot
\vec{\nabla}\nonumber\\ &-& \frac{\gamma^0}{2m}\,i\,\vec{\Sigma}\cdot
\Big(\vec{\nabla}(U_+ + 2U_-)\times \vec{\nabla}\,\Big) -
\frac{\gamma^0}{m}\,(1 + U_+ + 2 U_-)\Delta -
\frac{\gamma^0}{2m}\,\Delta (U_+ +
U_-)\nonumber\\ &+&\frac{1}{4}\,\gamma^5\,{\cal K} +
\frac{i}{2m}\,{\cal K}\,\gamma^0\,\vec{\Sigma}\cdot \vec{\nabla} +
\frac{i}{4m}\,\gamma^0\,\vec{\Sigma}\cdot \vec{\nabla}{\cal
  K}\nonumber\\ &+& \frac{1}{4m}\,\vec{\gamma}\cdot(\vec{\cal B}\times
\vec{\nabla}\,) - \frac{1}{8m}\,\vec{\gamma}\cdot {\rm rot}\vec{\cal
  B} + \frac{i}{8m}\,\gamma^0\,\gamma^5\,{\rm div}\vec{\cal
  B},\nonumber\\ \frac{i^2}{2}\,\Big[S_1,\Big[S_1,{\rm H}' -
    \frac{1}{3}\,\frac{\partial S_1}{\partial
      t}\Big]\Big]&=&\frac{\gamma^0}{2m}\,(1 + U_+ + 2 U_-)\,\Delta +
\frac{\gamma^0}{m}\,\vec{\nabla}(U_+ + U_-)\cdot \vec{\nabla} +
\frac{\gamma^0}{8m}\,\Delta(3 U_+ + 2 U_-)\nonumber\\ &+&
\frac{\gamma^0}{4m}\,i\,\vec{\Sigma}\cdot \Big(\vec{\nabla}(U_+ + 2
U_-) \times \vec{\nabla}\,\Big) - \frac{i}{4m}\,{\cal
  K}\,\gamma^0\,\vec{\Sigma}\cdot \vec{\nabla} -
\frac{i}{8m}\,\gamma^0\,\vec{\Sigma}\cdot \vec{\nabla}{\cal K}.
\end{eqnarray}
Thus, the unitary transformation Eq.(\ref{eq:32})
yields the following Hamilton operator of slow
fermions
\begin{eqnarray}\label{eq:35}
{\rm H}_1 &=& \gamma^0 m - \frac{\gamma^0}{2m}\,\Delta + \gamma^0 m
U_+ + \frac{i}{2}\,\gamma^0\vec{\gamma}\cdot \vec{\nabla}U_+ -
\frac{\gamma^0}{2m}\,\vec{\nabla}(U_+ + 2 U_-)\cdot \vec{\nabla} -
\frac{\gamma^0}{2m}\,(U_+ + 2 U_-)\,\Delta -
\frac{\gamma^0}{8m}\,\Delta(U_+ + 2U_-)\nonumber\\ &-&
\frac{\gamma^0}{4m}\,i\,\vec{\Sigma}\cdot \Big(\vec{\nabla}(U_+ + 2
U_-) \times \vec{\nabla}\,\Big) - \frac{1}{4}\,\vec{\Sigma}\cdot
\vec{\cal B} + \frac{i}{4m}\,{\cal K}\,\gamma^0\,\vec{\Sigma}\cdot
\vec{\nabla} + \frac{i}{8m}\,\gamma^0\,\vec{\Sigma}\cdot
\vec{\nabla}{\cal K} \nonumber\\ &+& \frac{1}{4m}\,\vec{\gamma}\cdot
(\vec{\cal B}\times \vec{\nabla}\,) - \frac{1}{8m}\,\vec{\gamma}\cdot
    {\rm rot}\vec{\cal B} + \frac{i}{8m}\,\gamma^0\,\gamma^5\,{\rm
      div}\vec{\cal B} + \frac{1}{2m}\,\frac{\partial U_-}{\partial
      t}\,\vec{\gamma}\cdot \vec{\nabla} - \frac{i}{8m}\,\gamma^0
    \gamma^5\,\frac{\partial {\cal K}}{\partial t},
\end{eqnarray}
where we have neglected the contributions of order
  $1/m^2$.  In order to remove {\it odd} operators in the Hamilton
operator ${\rm H}_1$ we perform the second unitary transformation
\cite{Foldy1950}
\begin{eqnarray}\label{eq:36}
{\rm H}_2 = e^{\,+ i S_2}\,{\rm H}_1\,e^{\,-i S_2} - i\,e^{\,i
  S_2}\frac{\partial}{\partial t}e^{\,-i S_2} = {\rm H}_1 -
\frac{\partial S_2}{\partial t} + i\Big[S_2,{\rm H}_1 -
  \frac{1}{2}\,\frac{\partial S_2}{\partial t}\Big] +
\frac{i^2}{2}\,\Big[S_2,\Big[S_2,{\rm H}_1 -
    \frac{1}{3}\,\frac{\partial S_2}{\partial t}\Big]\Big] + \ldots,
\end{eqnarray}
where the operator $S_2$ is equal to
\begin{eqnarray}\label{eq:37}
S_2 &=& - \frac{i}{2m}\,\gamma^0\Big(\frac{i}{2}\,\gamma^0
\vec{\gamma}\cdot \vec{\nabla}U_+ + \frac{1}{4m}\,\vec{\gamma}\cdot
(\vec{\cal B}\times \vec{\nabla}\,) - \frac{1}{8m}\,\vec{\gamma}\cdot
    {\rm rot}\vec{\cal B} + \frac{i}{8m}\,\gamma^0\,\gamma^5\,{\rm
      div}\vec{\cal B} + \frac{1}{2m}\,\frac{\partial U_-}{\partial
      t}\,\vec{\gamma}\cdot \vec{\nabla} - \frac{i}{8m}\,\gamma^0
    \gamma^5\,\frac{\partial {\cal K}}{\partial t}\Big) =
    \nonumber\\ &=& \frac{1}{4m}\,\vec{\gamma}\cdot \vec{\nabla}U_+ -
    \frac{i}{8m^2}\,\gamma^0\,\vec{\gamma}\cdot (\vec{\cal B}\times
    \vec{\nabla}\,) + \frac{i}{16m^2}\,\gamma^0\,\vec{\gamma}\cdot
        {\rm rot}\vec{\cal B} + \frac{1}{16m^2}\,\gamma^5\,{\rm
          div}\vec{\cal B} - \frac{1}{4m^2}\,\frac{\partial
          U_-}{\partial t}\,i\,\gamma^0\,\vec{\gamma}\cdot
        \vec{\nabla} - \frac{1}{16m^2}\,\gamma^5\,\frac{\partial
          {\cal K}}{\partial t}.\nonumber\\ 
&&
\end{eqnarray}
Keeping only the contributions to order $1/m$ for the time derivative
of $S_2$ and the commutators we obtain the following expressions
\begin{eqnarray}\label{eq:38}
\frac{\partial S_2}{\partial t} &=& \frac{1}{4m}\,\vec{\gamma}\cdot
\vec{\nabla}\frac{\partial U_+}{\partial t},\nonumber\\ i\Big[S_2,{\rm
    H}_1 - \frac{1}{2}\,\frac{\partial S_2}{\partial t}\Big]&=& -
\frac{i}{2}\,\gamma^0\vec{\gamma}\cdot \vec{\nabla}U_+ -
\frac{1}{4m}\,\vec{\gamma}\cdot (\vec{\cal B}\times \vec{\nabla}\,) +
\frac{1}{8m}\,\vec{\gamma}\cdot {\rm rot}\vec{\cal B} -
\frac{i}{8m}\,\gamma^0\,\gamma^5\,{\rm div}\vec{\cal B},
\nonumber\\ \frac{i^2}{2}\,\Big[S_2,\Big[S_2,{\rm H}_1 -
    \frac{1}{3}\,\frac{\partial S_2}{\partial t}\Big]\Big]&=& 0.
\end{eqnarray}
Thus, after two unitary transformation the effective Hamilton operator
of slow fermions takes the form
\begin{eqnarray}\label{eq:39}
{\rm H}_2 &=& \gamma^0 m - \frac{\gamma^0}{2m}\,\Delta\nonumber\\ &+&
\gamma^0 m U_+ - \frac{\gamma^0}{2m}\,\vec{\nabla}(U_+ + 2 U_-)\cdot
\vec{\nabla} - \frac{\gamma^0}{2m}\,(U_+ + 2 U_-)\,\Delta -
\frac{\gamma^0}{8m}\,\Delta(U_+ + 2U_-) -
\frac{\gamma^0}{4m}\,i\,\vec{\Sigma}\cdot \Big(\vec{\nabla}(U_+ + 2
U_-) \times \vec{\nabla}\,\Big)\nonumber\\ &-&
\frac{1}{4}\,\vec{\Sigma}\cdot \vec{\cal B} + \frac{i}{4m}\,{\cal
  K}\,\gamma^0\,\vec{\Sigma}\cdot \vec{\nabla} +
\frac{i}{8m}\,\gamma^0\,\vec{\Sigma}\cdot \vec{\nabla}{\cal K} -
\frac{1}{4m}\,\vec{\gamma}\cdot \vec{\nabla}\frac{\partial
  U_+}{\partial t}.
\end{eqnarray}
The last term in Eq.(\ref{eq:39}), which is the last
  remained {\it odd} operator after two FW transformations, we delete
by the third FW transformation
\begin{eqnarray}\label{eq:40}
{\rm H}_3 = e^{\,+ i S_3}\,{\rm H}_2\,e^{\,-i S_3} - i\,e^{\,i
  S_3}\frac{\partial}{\partial t}e^{\,-i S_3} = {\rm H}_2 -
\frac{\partial S_3}{\partial t} + i\Big[S_3,{\rm H}_3 -
  \frac{1}{2}\,\frac{\partial S_3}{\partial t}\Big] +
\frac{i^2}{2}\,\Big[S_3,\Big[S_3,{\rm H}_1 -
    \frac{1}{3}\,\frac{\partial S_3}{\partial t}\Big]\Big] + \ldots
\end{eqnarray}
with the operator $S_3$, given by
\begin{eqnarray}\label{eq:41}
S_3 = - \frac{i}{2m}\,\gamma^0\,\Big(- \frac{1}{4m}\,\vec{\gamma}\cdot
\vec{\nabla}\frac{\partial U_+}{\partial t}\Big) =
\frac{i}{8m^2}\,\gamma^0\,\vec{\gamma}\cdot \vec{\nabla}\frac{\partial
  U_+}{\partial t}.
\end{eqnarray}
Neglecting the contributions of the terms of order $1/m^2$ we obtain
the low--energy reduction of the Dirac Hamilton operator for slow
fermions
\begin{eqnarray}\label{eq:42}
{\rm H}_3 &=& \gamma^0 m - \frac{\gamma^0}{2m}\,\Delta\nonumber\\ &+&
\gamma^0 m U_+ - \frac{\gamma^0}{2m}\,\vec{\nabla}(U_+ + 2 U_-)\cdot
\vec{\nabla} - \frac{\gamma^0}{2m}\,(U_+ + 2 U_-)\,\Delta -
\frac{\gamma^0}{8m}\,\Delta(U_+ + 2U_-) -
\frac{\gamma^0}{4m}\,i\,\vec{\Sigma}\cdot \Big(\vec{\nabla}(U_+ + 2
U_-) \times \vec{\nabla}\,\Big)\nonumber\\ &-&
\frac{1}{4}\,\vec{\Sigma}\cdot \vec{\cal B} + \frac{i}{4m}\,{\cal
  K}\,\gamma^0\,\vec{\Sigma}\cdot \vec{\nabla} +
\frac{i}{8m}\,\gamma^0\,\vec{\Sigma}\cdot \vec{\nabla}{\cal K}.
\end{eqnarray}
Following the standard procedure \cite{Foldy1950}, removing the mass
term $\gamma^0 m$ and skipping intermediate calculations we arrive at
the Schr\"odinger--Pauli equation for the large component
$\Psi(\vec{r},t)$ of the Dirac wave function of slow
fermions
\begin{eqnarray}\label{eq:43}
i\frac{\partial \Psi(\vec{r},t)}{\partial t} = \Big( -
\frac{1}{2m}\,\Delta + m U_{\rm E} + \Phi_{\rm ngr-ch} + \Phi_{\rm
  mct}\Big)\,\Psi(\vec{r},t),
\end{eqnarray}
where $\Phi_{\rm ngr-ch}$ and $\Phi_{\rm mt}$ are the effective
low--energy gravitational potentials 
\begin{eqnarray}\label{eq:44}
\Phi_{\rm ngr-ch} &=& m (U_+ - U_{\rm E}) - \frac{1}{2m}\,\vec{\nabla}(U_+ + 2
U_-)\cdot \vec{\nabla} - \frac{1}{2m}\,(U_+ + 2 U_-)\,\Delta -
\frac{1}{8m}\,\Delta(U_+ + 2U_-)\nonumber\\ &-&
\frac{i}{4m}\,\vec{\sigma}\cdot \Big(\vec{\nabla}(U_+ + 2 U_-) \times
\vec{\nabla}\,\Big)
\end{eqnarray}
and 
\begin{eqnarray}\label{eq:45}
\Phi_{\rm mt} = -
\frac{1}{4}\,\vec{\sigma}\cdot \vec{\cal B} + \frac{i}{4m}\,{\cal
  K}\,\vec{\sigma}\cdot \vec{\nabla} + \frac{i}{8m}\,\vec{\sigma}\cdot
\vec{\nabla}{\cal K},
\end{eqnarray}
describing interactions of slow fermions with the chameleon and
torsion field, respectively, in the spacetime with the Schwarzschild
metric taken in the weak gravitational field approximation. The
contributions of the potentials $\Phi_{\rm ngr-ch}$ and $\Phi_{\rm
  mct}$ provide a deviation from the Newtonian potential of the Earth
$m U_{\rm E}$. The abbreviation (ngr-ch) means the {\it Newtonian
  gravitational potential with chameleon} potential, whereas the
abbreviation (mt) stands for the {\it minimal torsion} coupling.

After the replacement $U_+ \to U$ and $U_- \to \gamma\,U$ the
potential $\Phi_{\rm ngr-ch}$ in Eq.(\ref{eq:44}) coincides with the
effective low--energy gravitational potential Eq.(\ref{eq:1}),
calculated in \cite{Ivanov2014} (see also Eq.(\ref{eq:20}) of
Ref.\cite{Ivanov2014}). The difference, appearing in the effective
coupling constants, can be explained as follows. The chameleon field
in \cite{Ivanov2014} has been added to the weak gravitational field in
the form Eq.(\ref{eq:2}), whereas in this paper we have used a
relation $\tilde{g}_{\mu\nu} = f^2 g_{\mu\nu}$, connecting the metric
tensor $\tilde{g}_{\mu\nu}$ in the Jordan frame with the metric tensor
$g_{\mu\nu}$ in the Einstein frame through the conformal factor $f =
e^{\,\beta\phi/M_{\rm Pl}}$ \cite{Chameleon1,Chameleon2}. This leads
to the contributions of the chameleon field to the metric tensor
$\tilde{g}_{\mu\nu}$ as it is shown in Eq.(\ref{eq:21}).

 The first two terms in the effective low--energy gravitational
 potential $\Phi_{\rm mt}$, given by Eq.(\ref{eq:45}), were obtained
 by L\"ammerzahl \cite{Laemmerzahl1997}, where the derivatives of the
 torsion field were neglected. In turn, Obukhov, Silenko, and Teryaev
 \cite{Obukhov2014} kept the derivatives of the torsion field and the
 all terms of the effective low--energy gravitational potential
 $\Phi_{\rm mt}$ can be found in the effective Hamilton operator
 ${\cal H}^{(3)}_{\rm FW}$ of Ref.\cite{Obukhov2014} by expanding this
 operator in powers of $1/m$.

\section{Non--minimal coupling of torsion with slow fermions}
\label{sec:torsion}

The torsion tensor field ${\cal T}_{\sigma\mu\nu}$, being a
third--order tensor antisymmetric with respect to $\mu$ and $\nu$
indices ${\cal T}_{\sigma\mu\nu} = - {\cal T}_{\sigma\nu\mu}$, is
defined by $24$ independent components. They can be represented in the
following irreducible form \cite{Kostelecky2008}
\begin{eqnarray}\label{eq:46}
{\cal T}_{\sigma\mu\nu} = \frac{1}{3}\,(g_{\sigma\mu}{\cal E}_{\nu} -
g_{\sigma\nu}{\cal E}_{\mu}) + 
\frac{1}{3}\,\epsilon_{\sigma\mu\nu\beta}\, {\cal B}^{\beta} + {\cal
  M}_{\sigma\mu\nu},
\end{eqnarray}
where the 4--vector and axial 4--vector fields ${\cal E}_{\nu}$ and
${\cal B}^{\alpha}$ fields, respectively, possessing 4 independent
components each, are defined by
\begin{eqnarray}\label{eq:47}
{\cal E}_{\nu} = g^{\sigma\mu}{\cal T}_{\sigma\mu\nu}\quad,\quad {\cal B}^{\alpha} =
 \frac{1}{2}\,\epsilon^{\alpha\sigma\mu\nu}{\cal
  T}_{\sigma\mu\nu}.
\end{eqnarray}
For the definition of the vector field ${\cal B}^{\alpha}$ in terms of
the torsion tensor field ${\cal T}_{\sigma\mu\nu}$ we have used the
relation $\epsilon^{\alpha\sigma\mu\nu}\epsilon_{\sigma\mu\nu\beta} =
- 6\,{\delta^{\alpha}}_{\beta}$ \cite{Itzykson1980}. The residual 16
independent components can be attributed to the tensor field ${\cal
  M}_{\sigma\mu\nu}$, which obeys the constraints $g^{\sigma\mu}{\cal
  M}_{\sigma\mu\nu} = \epsilon^{\alpha\sigma\mu\nu}{\cal
  M}_{\sigma\mu\nu} = 0$ \cite{Kostelecky2008}.

As we have shown above the effective low--energy potential of slow
fermions, coupled minimally to the torsion, contains only the
axial 4--vector torsion components ${\cal
  B}^{\alpha}$. The 4-vector ${T^{\alpha}}_{\alpha\mu} = {\cal
  E}_{\mu}$ and tensor ${\cal M}_{\sigma\mu\nu}$ torsion components
have no minimal couplings with Dirac fermions.  The most general
phenomenological Lagrangian of Dirac fermions coupled
  to torsion has been proposed by Kostelecky, Russell, and Tasson
  \cite{Kostelecky2008}. We would like to mention that the
  non--minimal torsion--matter couplings have been recently discussed
  by Puetzfeld and Obukhov \cite{Obukhov2014a}.

In this section we derive the effective low--energy potential for slow
fermions, coupled to the 4--vector ${\cal E}_{\mu}$ and tensor ${\cal
  M}_{\sigma\mu\nu}$ torsion components. Following Kostelecky, Russell,
and Tasson \cite{Kostelecky2008} we consider the
  torsion--fermion Lagrangian
\begin{eqnarray}\label{eq:48}
 \delta {\cal L}_{\rm nmt} &=& \frac{1}{4}\,\kappa_1\,\sqrt{-
   \tilde{g}}\,{\tilde{\cal T}^{\alpha}\,\!\!}_{\alpha\mu}(x)\,
 \tilde{e}^{\mu}_{\hat{\lambda}}(x)
 \,\bar{\psi}(x)\gamma^{\hat{\lambda}}\psi(x)\nonumber\\ &+&
 \frac{\kappa_2}{8m}\,\sqrt{- \tilde{g}}\,i\,{\tilde{\cal
     T}^{\alpha}\,\!\!}_{\alpha\mu}(x)\,\tilde{e}^{\mu}_{\hat{\alpha}}(x)\,
 \tilde{e}^{\nu}_{\hat{\beta}}(x)\,\Big(\bar{\psi}(x)
 \sigma^{\hat{\alpha}\hat{\beta}} D_{\nu}\psi(x) -
 (\bar{\psi}(x)\bar{D}_{\nu})\sigma^{\hat{\alpha}\hat{\beta}}\psi(x)\Big)
 \nonumber\\ &+& \frac{\kappa_3}{8 m}\,\sqrt{-
   \tilde{g}}\,i\,{\tilde{\cal
     M}^{\lambda}\,\!\!}_{\mu\nu}(x)\,\tilde{e}^{\mu}_{\hat{\alpha}}(x)\,
 \tilde{e}^{\nu}_{\hat{\beta}}(x)\,\Big(\bar{\psi}(x)
 \sigma^{\hat{\alpha}\hat{\beta}} D_{\lambda}\psi(x) -
 (\bar{\psi}(x)\bar{D}_{\lambda})\sigma^{\hat{\alpha}\hat{\beta}}\psi(x)\Big),
 \end{eqnarray}
where $\kappa_j$ $(j = 1,2,3)$ are phenomenological dimensionless
coupling constants and the abbreviation (nmt) means the {\it
  non--minimal torsion} coupling. The phenomenological Lagrangian
Eq.(\ref{eq:48}) of non--minimal torsion couplings is a generalization
of the phenomenological Lagrangian, proposed by Kostelecky, Russell,
and Tasson \cite{Kostelecky2008}, on a curved spacetime with the
chameleon field. We would like to note that from the general
phenomenological Lagrangian, proposed in \cite{Kostelecky2008}, we
have taken only the interactions, which are invariant under parity
(P), charge--parity (CP) and time--reversal (T) transformations, and
neglected the contribution of the non--derivative axial 4--vector
torsion--fermion interaction, since it is already
taken into account as the minimal torsion--fermion coupling.

To linear order approximation of interacting fields the Hamilton
operator, corresponding to the phenomenological Lagrangian
Eq.(\ref{eq:48}), takes the form
\begin{eqnarray}\label{eq:49}
\delta {\rm H}_{\rm nmt} &=& - \frac{1}{4}\,\kappa_1\,{\cal
  E}_{\mu}\gamma^0\gamma^{\mu} - \frac{\kappa_2}{4 m}\,i\,{\cal
  E}_{\mu}\gamma^0 \sigma^{\mu\nu}\partial_{\nu} - \frac{\kappa_2}{8
  m}\,i\,\partial_{\nu}{\cal E}_{\mu}\gamma^0
\sigma^{\mu\nu}\nonumber\\ &&- \frac{\kappa_3}{4 m}\,i\,{{\cal
    M}^{\lambda}}_{\mu\nu}\gamma^0 \sigma^{\mu\nu}\partial_{\lambda} -
\frac{\kappa_3}{8 m}\,i\,\partial_{\lambda}{{\cal
    M}^{\lambda}}_{\mu\nu}\gamma^0 \sigma^{\mu\nu}.
\end{eqnarray}
Using the relations for the Dirac matrices
\begin{eqnarray}\label{eq:50}
 \gamma^0\sigma^{\mu\nu} &=& i(\eta^{0\mu}\gamma^{\nu} -
 \eta^{0\nu}\gamma^{\mu}) -
 \epsilon^{0\mu\nu\rho}\gamma_{\rho}\gamma^5,\nonumber\\ \gamma^0
 \gamma^j\gamma^k &=& \gamma^0\,\eta^{jk} +
 i\,\epsilon^{jk\ell}\gamma^0 \Sigma_{\ell},
\end{eqnarray}
where $\Sigma_{\ell} = \gamma^0 \gamma_{\ell} \gamma^5$, we arrive at
the Hamilton operator of the non--minimal torsion--fermion (matter)
couplings
\begin{eqnarray}\label{eq:51}
 \delta {\rm H}_{\rm nmt} &=& - \kappa_1\,\frac{1}{4}\,{\cal E}_0 +
 \kappa_1\,\frac{1}{4}\,\gamma^0 \vec{\gamma}\cdot\vec{\cal E} +
 \frac{\kappa_2}{4m}\,\vec{\gamma}\cdot \vec{\cal
   E}\,\frac{\partial}{\partial t} +
 \frac{\kappa_2}{8m}\,\vec{\gamma}\cdot \frac{\partial \vec{\cal
     E}}{\partial t} + \frac{\kappa_2}{4m}\,{\cal
   E}_0\,\vec{\gamma}\cdot \vec{\nabla} +
 \frac{\kappa_2}{8m}\,\vec{\gamma}\cdot \vec{\nabla}{\cal E}_0
 \nonumber\\ && + \frac{\kappa_2}{4m}\,i\,\gamma^0
 \vec{\Sigma}\cdot(\vec{\cal E}\times \vec{\nabla}\,) -
 \frac{\kappa_2}{8m}\,i\,\gamma^0 \vec{\Sigma}\cdot {\rm
   rot}\,\vec{\cal E} + \frac{\kappa_3}{2m}\,{{\cal
     M}^{\lambda}}_{0j}\gamma^j\partial_{\lambda} +
 \frac{\kappa_3}{4m}\,\partial_{\lambda}{{\cal M}^{\lambda}}_{0j}\,\gamma^j\nonumber\\
&& + i\,\frac{\kappa_3}{4m}\,\epsilon^{jk\ell}\,\gamma^0 \Sigma_{\ell}\,{{\cal M}^{\lambda}}_{jk}\,\partial_{\lambda} +  i\,\frac{\kappa_3}{4m}\,\epsilon^{jk\ell}\,\gamma^0 \Sigma_{\ell}\,\partial_{\lambda}{{\cal M}^{\lambda}}_{jk},
\end{eqnarray}
where we have used that ${{\cal T}^{\alpha}}_{\alpha j} = ( -
\vec{\cal E})_j$.  Skipping standard intermediate Foldy--Wouthuysen
calculations we arrive at the effective low--energy gravitational
potential for slow fermions, non--minimally coupled to torsion:
\begin{eqnarray}\label{eq:52}
 \Phi_{\rm nmt} &=& - \kappa_1\,\frac{1}{4}\,{\cal E}_0 -
 i\,\frac{\kappa_1}{2m}\,\vec{\cal E}\cdot \vec{\nabla} -
 i\,\frac{\kappa_1}{4m}\,{\rm div}\,\vec{\cal E} +
 \frac{\kappa_1}{4m}\,\vec{\sigma}\cdot {\rm rot}\,\vec{\cal
   E}\nonumber\\ &&+ i\,\frac{\kappa_2}{4m}\,\vec{\sigma}\cdot
 (\vec{\cal E}\times \vec{\nabla}\,) -
 i\,\frac{\kappa_2}{8m}\,\vec{\sigma}\cdot {\rm rot}\,\vec{\cal
   E}\nonumber\\ &&+
 \frac{1}{4}\,\kappa_3\,\epsilon^{jk\ell}\,\sigma_{\ell}\,{{\cal
     M}^0}_{jk} +
 i\,\frac{\kappa_3}{4m}\,\epsilon^{jk\ell}\,\sigma_{\ell}\,{{\cal
     M}^{\lambda}}_{jk}\partial_{\lambda} +
 i\,\frac{\kappa_3}{8m}\,\epsilon^{jk\ell}\,\sigma_{\ell}\,
 \partial_{\lambda}{{\cal M}^{\lambda}}_{jk},
\end{eqnarray}
where we have neglected the contributions of order $O(1/m^2)$ and
denoted $\sigma_{\ell} = (- \vec{\sigma}\,)_{\ell}$.

The important peculiarity of the effective low--energy potential
Eq.(\ref{eq:52}) is the appearance of non--spin--torsion--fermion
interactions with the coupling constants proportional to
$\kappa_1$. This agrees well with the observation, obtained by
Puetzfeld and Obukhov \cite{Obukhov2014a}. In turn, the
torsion--fermion interactions, proportional to the coupling constants
$\kappa_2$ and $\kappa_3$, are only spin--torsion--fermion ones.

The interactions, proportional to $\kappa_2$, do not contradict the
definition of the vector torsion components in terms of the scalar
field $\vec{\cal E} \sim \vec{\nabla}\,\varphi$, where $\varphi$ is a
scalar field. Since in this case ${\rm rot}\,\vec{\cal E} = 0$, the
spin --chameleon--matter interaction in Eq.(\ref{eq:44}) becomes
equivalent to the phenomenological non--minimal torsion--matter
derivative coupling of the vector torsion components in
Eq.(\ref{eq:48}). This agrees well the analysis of the 4--vector
torsion components, carried out in \cite{Ivanov2014,Ivanov2015}. The
tensor torsion components, described by ${{\cal
    M}^{\lambda}}_{\mu\nu}$, possess only spin--torsion--matter
couplings.

Thus, the Schr\"odinger--Pauli equation of slow fermions, coupled to
the gravitational field of the Earth, the chameleon field and torsion,
is given by
\begin{eqnarray}\label{eq:54}
i\frac{\partial \Psi(\vec{r},t)}{\partial t} = \Big( -
\frac{1}{2m}\,\Delta + m U_{\rm E} + \Phi_{\rm ngr-ch} + \Phi_{\rm
  mct} + \Phi_{\rm nmt}\Big)\,\Psi(\vec{r},t).
\end{eqnarray}
This equation can be, in principle, used for the analysis of the
fine--structure and transition frequencies of quantum gravitational
states of ultracold neutrons \cite{Abele2010}--\cite{AbeleWF3}.

\section{Conclusive discussion}
\label{sec:conclusion}

We have analysed the low--energy approximation of the Dirac equation
for slow fermions, coupled to the chameleon field and torsion in the
spacetime with the Schwarzschild metric taken in the weak Newtonian
gravitational field of the Earth approximation. The aim of this
analysis is addressed to the derivation of an effective low--energy
gravitational potential for slow fermions coupled to gravitational and
chameleon fields and torsion with minimal and non--minimal
couplings. The obtained effective low--energy gravitational potential
$\Phi_{\rm eff} = \Phi_{\rm ngr-ch} + \Phi_{\rm mt} + \Phi_{\rm nmt}$
for slow fermions (neutrons), coupled to the gravitational field of
the Earth, the chameleon field and torsion with minimal and
non--minimal torsion--fermion couplings, can be, in principle,
investigated experimentally in the terrestrial laboratories in the
qBounce experiments \cite{Abele2010,Jenke2011} (see also
\cite{Jenke2014}), the quantum ball experiments
\cite{AbeleWF1,AbeleWF2,AbeleWF3} and neutron interferometry
\cite{Rauch2000,Lemmel2015}.

We have reproduced the main structure of the effective low--energy
gravitational potential for slow fermions in the weak gravitational
field of the Earth and the chameleon field, which was calculated in
\cite{Ivanov2014}. The distinction of our potential Eq.(\ref{eq:44})
from the potential Eq.(\ref{eq:20}) of Ref.\cite{Ivanov2014} is in the
coupling constants of the interactions of order $1/m$. Describing the
chameleon--fermion interactions in terms of the Jordan--frame metric
we become that the time--time $\tilde{g}_{00}$ and space--space
$\tilde{g}_{ij}$ components of the metric tensor are defined in terms
of the potentials $U_+ = U_{\rm E} + (\beta/M_{\rm Pl})\phi$ and $U_-
= U_{\rm E} - (\beta/M_{\rm Pl})\phi$, respectively. This diminishes
the chameleon--matter coupling constants of order $1/m$ by a factor of
3 and changes their signs in comparison with those in Eq.(\ref{eq:20})
of Ref.\cite{Ivanov2014} (see also Eq.(\ref{eq:1}) in section
\ref{sec:introduction}).  However, this does not change the assertion
that the torsion field can be a gradient of the chameleon one, which
has been recently developed in \cite{Ivanov2015} as a version of the
Einstein--Cartan gravity with torsion.

The effective low--energy torsion--fermion potential Eq.(\ref{eq:45})
agrees well with the effective low--energy torsion--fermion
potentials, derived by L\"ammerzahl \cite{Laemmerzahl1997} at the
neglect of the derivatives of the torsion field (see Eq.(22) of
Ref.\cite{Laemmerzahl1997}), and by Obukhov, Silenko, and Teryaev
\cite{Obukhov2014} by expanding the effective
Hamilton operator ${\cal H}^{(3)}_{\rm FW}$ in powers of $1/m$ to
order $1/m$.

In addition to the effective low--energy potential Eq.(\ref{eq:45}),
caused by the minimal torsion--fermion couplings, we have derived the
effective low--energy gravitational potential $\Phi_{\rm nmt}$ of the
non--minimal torsion--fermion couplings. An interesting peculiarity of
this effective low--energy potential $\Phi_{\rm nmt}$ is an appearance
of some non--spin--torsion--fermion couplings, caused by non--minimal
non--derivative couplings of vector torsion
components. This agrees well with the results,
obtained by Puetzfeld and Obukhov \cite{Obukhov2014a}.  In turn, in
the low--energy approximation the phenomenological derivative
non--minimal torsion--fermion interactions of vector ${\cal E}_{\mu}$
and tensor ${{\cal M}^{\lambda}}_{\mu\nu}$ torsion components possess
only spin--torsion--fermion couplings. Therewith the derivative
non--minimal torsion--fermion couplings of vector torsion components
agree well with the hypothesis, proposed in \cite{Ivanov2014} and
developed in \cite{Ivanov2015}, that the vector torsion components can
be induced by the chameleon field. In this case the
spin--chameleon--fermion potential in Eq.(\ref{eq:44}) can be treated
as a low--energy approximation of the corresponding phenomenological
derivative torsion--fermion interaction of vector torsion components
in Eq.(\ref{eq:48}). As regards the torsion tensor components ${{\cal
    M}^{\lambda}}_{\mu\nu}$, we have to note that, according
Eq.(\ref{eq:52}), at low--energies only time--space--space ${{\cal
    M}^0}_{jk}$ and space--space--space ${{\cal M}^i}_{jk}$ torsion
components can be, in principle, observable.

The experimental upper bound of the spin--torsion--neutron coupling
has been recently obtained by Lehnert, Snow, and Yan
\cite{Lehnert2014}. By measuring a neutron spin rotation in the liquid
${^4}{\rm He}$, it has been found that a linear superposition $\zeta$
of vector and axial--vector torsion components, defined by Kostelecky
{\it et al.}  \cite{Kostelecky2008}, is restricted from above $|\zeta|
< 5.4\times 10^{-16}\,{\rm GeV}$ at $68\,\%$ of C.L..

The numerical estimates of the upper bounds of the axial--vector
torsion components have been carried out by L\"ammerzahl
\cite{Laemmerzahl1997}, Kostelecky, Russell, and Tasson
\cite{Kostelecky2008}, and Obukhov, Silenko, and Teryaev
\cite{Obukhov2014}. In turn, Kostelecky {\it et al.}
\cite{Kostelecky2008} have given also the estimates of the vector and
tensor torsion components. The results, obtained by Kostelecky {\it et
  al.} \cite{Kostelecky2008} for the axial--vector torsion components
in the minimal torsion--matter coupling approach, are: $|{\cal B}_T| <
1.0\times 10^{-27}\,{\rm GeV}$, $|{\cal B}_X| < 7.0\times
10^{-32}\,{\rm GeV}$, $|{\cal B}_Y| < 8.4 \times 10^{-32}\,{\rm GeV}$
and $|{\cal B}_Z| < 3.4\times 10^{-30}\,{\rm GeV}$ (see
Eq.(\ref{eq:6}) of Ref.\cite{Kostelecky2008}), where we have used the
relation $\vec{{\cal B}} = 3\,\vec{A}$. They agree well with the
estimates, given by L\"ammerzahl \cite{Laemmerzahl1997} and by
Obukhov, Silenko, and Teryaev \cite{Obukhov2014}.

The estimates of the vector $\xi^{(a)}_j{\cal E}_{\mu}$ and and tensor
$\xi^{(a)}_j{\cal M}_{\lambda\mu\nu}$ torsion components from
derivative torsion--fermion couplings, multiplied by the corresponding
phenomenological coupling constants $\xi^{(a)}_j$, vary over the range
$(10^{-31} - 10^{-26})$. For example, the upper bound of the product
$|\xi^{(5)}_5 {\cal M}_{\alpha\mu\nu}|$ varies for different
components of the torsion tensor components ${\cal M}_{\alpha\mu\nu}$
from $10^{-31}$ to $10^{-26}$ (see Table I of
Ref.\cite{Kostelecky2008}). The largest value $10^{-26}$ appears for
the time--space--space torsion tensor component $|\xi^{(5)}_5 {\cal
  M}_{TXY}| < 10^{-26}$.  It is important to note that estimates,
carried out in \cite{Kostelecky2008}, have been done for constant
torsion components. This has led to the disappearance of the products
$\xi^{(4)}_1{\cal E}_{\mu}$ and $\xi^{(5)}_6 {\cal E}_{\mu}$, where
coupling constants $\xi^{(4)}_1$ and $\xi^{(5)}_6$ are equal to
$\xi^{(4)}_1 = \kappa_1/4$ and $\xi^{(5)}_6 = \kappa_2/4m$,
respectively, from the superpositions of torsion--matter couplings the
upper bounds for which can be obtained from the experiments on the
verification of the Standard Model Extension (SME)
\cite{KosteleckyData}. This gives a chance that the contributions of
these torsion--fermion interactions can be estimated from the
experimental data in the terrestrial laboratories, where a space--time
dependence of torsion is taken into account. The experimental upper
bound $|\zeta| < 5.4\times 10^{-16}\,{\rm GeV}$, reported by Lehnert,
Snow, and Yan \cite{Lehnert2014}, is by twelve orders of magnitude
larger compared to the estimate $|\zeta| < 10^{-27}\,{\rm GeV}$, which
can be obtained from Table I of Ref.\cite{Kostelecky2008}. Completing
our discussion we would like to note that we have considered the
low--energy approximation of a part of the phenomenological
relativistic invariant torsion--fermion interactions only. The
analysis of the low--energy approximation of the rest of the
phenomenological torsion--fermion interactions, proposed by
Kostelecky, Russell, and Tasson \cite{Kostelecky2008}, we are planning
to perform in our forthcoming publication.

\section{Acknowledgements}

We are grateful to Hartmut Abele for stimulating discussions and Alan
Kostelecky for fruitful discussions of the properties of the Dirac
fermions in the curved spacetime with torsion. This work was supported
by the Austrian ``Fonds zur F\"orderung der Wissenschaftlichen
Forschung'' (FWF) under the contract I689-N16.

\section{Appendix A: Covariant derivation of the Dirac equation 
in the curved spacetime with the chameleon field and torsion}
\renewcommand{\theequation}{A-\arabic{equation}}
\setcounter{equation}{0}

For the investigation of the dynamics of fermions, coupled to the
gravitational field with torsion and the chameleon field through the
Jordan metric tensor $\tilde{g}_{\mu\nu}$, we follow Kostelecky
\cite{Kostelecky2004} and define by the action
\begin{eqnarray}\label{eq:A.1}
{\rm S}_{\psi} = \int d^4x\,\sqrt{-
  \tilde{g}}\,\Big(i\,\frac{1}{2}\,\bar{\psi}(x)\tilde{\gamma}^{\mu}(x)
\stackrel{\leftrightarrow}{D}_{\mu}\psi(x) - m\bar{\psi}(x)\psi(x)\Big).
\end{eqnarray}
The definition $\bar{\psi}(x)
\tilde{\gamma}^{\mu}(x)\!\!\stackrel{\leftrightarrow}{D}_{\mu}\!\!\psi(x)$
should be understood as follows \cite{Kostelecky2004}
\begin{eqnarray}\label{eq:A.2}
 \bar{\psi}(x)\,
 \tilde{\gamma}^{\mu}(x)\!\stackrel{\leftrightarrow}{D}_{\mu}\!\psi(x) =
 \tilde{e}^{\mu}_{\hat{\alpha}}(x)\Big(
 \bar{\psi}(x)\gamma^{\hat{\alpha}}D_{\mu}\psi(x) -
 (\bar{\psi}\bar{D}_{\mu})\gamma^{\hat{\alpha}}\psi(x)\Big),
\end{eqnarray}
where $(\bar{\psi}\bar{D}_{\mu})$ means 
\begin{eqnarray}\label{eq:A.3}
 (\bar{\psi}\bar{D}_{\mu}) = \partial_{\mu}\bar{\psi}(x) -
  \bar{\psi}(x)
  \gamma^{\hat{0}}\tilde{\Gamma}^{\dagger}_{\mu}(x)\gamma^{\hat{0}}.
\end{eqnarray}
For the derivation of the covariant Dirac equation in the curved
spacetime we rewrite the action Eq.(\ref{eq:A.1}) in terms of the
vierbein fields and the spin connection. We get
\begin{eqnarray}\label{eq:A.4}
{\rm S}_{\psi} &=& \int d^4x\,\sqrt{-
  \tilde{g}}\,\Big(i\,\frac{1}{2}\,\tilde{e}^{\mu}_{\hat{\lambda}}(x)
\bar{\psi}(x) \gamma^{\hat{\lambda}} \Big(\partial_{\mu}\psi(x) -
\frac{1}{4}\,i\,\tilde{\omega}_{\mu \hat{\alpha}\hat{\beta}}(x)
\sigma^{\hat{\alpha}\hat{\beta}}\psi(x)\Big)\nonumber\\ &-&
i\,\frac{1}{2}\,\tilde{e}^{\mu}_{\hat{\lambda}}(x)
\Big(\partial_{\mu}\bar{\psi}(x) + \frac{1}{4}\,i\,\tilde{\omega}_{\mu
  \hat{\alpha}\hat{\beta}}(x) \bar{\psi}(x)
\sigma^{\hat{\alpha}\hat{\beta}}\Big) \gamma^{\hat{\lambda}} \psi(x) -
m\,\bar{\psi}(x)\psi(x)\Big).
\end{eqnarray}
Thus, the Lagrangian of the fermion in the curved spacetime is equal
to
\begin{eqnarray}\label{eq:A.5}
{\cal L}_{\psi} &=& \sqrt{-
  \tilde{g}}\,\Big(i\,\frac{1}{2}\,\tilde{e}^{\mu}_{\hat{\lambda}}(x)
\bar{\psi}(x) \gamma^{\hat{\lambda}} \Big(\partial_{\mu}\psi(x) -
\frac{1}{4}\,i\,\tilde{\omega}_{\mu \hat{\alpha}\hat{\beta}}(x)
\sigma^{\hat{\alpha}\hat{\beta}}\psi(x)\Big)\nonumber\\ &-&
i\,\frac{1}{2}\,\tilde{e}^{\mu}_{\hat{\lambda}}(x)
\Big(\partial_{\mu}\bar{\psi}(x) + \frac{1}{4}\,i\,\tilde{\omega}_{\mu
  \hat{\alpha}\hat{\beta}}(x) \bar{\psi}(x)
\sigma^{\hat{\alpha}\hat{\beta}}\Big) \gamma^{\hat{\lambda}} \psi(x) -
m\,\bar{\psi}(x)\psi(x)\Big).
\end{eqnarray}
The equation of motion of the fermion in the curved spacetime or the
Dirac equation is
\begin{eqnarray}\label{eq:A.6}
\partial_{\mu}\frac{\delta {\cal L}_{\psi}}{\delta
    \partial_{\mu}\bar{\psi}} = \frac{\delta {\cal L}_{\psi}}{\delta
  \bar{\psi}},
\end{eqnarray}
where
\begin{eqnarray}\label{eq:A.7}
\frac{\delta {\cal L}_{\psi}}{\delta \partial_{\mu}\bar{\psi}}&=& -
\sqrt{- \tilde{g}}\,i\,\frac{1}{2}\,\tilde{e}^{\mu}_{\hat{\lambda}}(x)
\gamma^{\hat{\lambda}} \psi(x),\nonumber\\ \frac{\delta {\cal
    L}_{\psi}}{\delta \bar{\psi}}&=& \sqrt{-
  \tilde{g}}\,i\,\frac{1}{2}\,\tilde{e}^{\mu}_{\hat{\lambda}}(x)
\gamma^{\hat{\lambda}} \Big(\partial_{\mu}\psi(x) -
\frac{1}{4}\,i\,\tilde{\omega}_{\mu \hat{\alpha}\hat{\beta}}(x)
\sigma^{\hat{\alpha}\hat{\beta}}\psi(x)\Big)\nonumber\\ &+& \sqrt{-
  \tilde{g}}\,\frac{1}{8}\,\tilde{e}^{\mu}_{\hat{\lambda}}(x)\,\tilde{\omega}_{\mu
  \hat{\alpha}\hat{\beta}}(x) \sigma^{\hat{\alpha}\hat{\beta}}
\gamma^{\hat{\lambda}} \psi(x) - \sqrt{- \tilde{g}}\,m\,\psi(x).
\end{eqnarray}
Substituting Eq.(\ref{eq:A.7}) into Eq.(\ref{eq:A.6}) we obtain the
Dirac equation in the following form
\begin{eqnarray}\label{eq:A.8}
\Big( i\,\tilde{e}^{\mu}_{\hat{\lambda}}(x) \gamma^{\hat{\lambda}}
D_{\mu} + \frac{1}{2}\,i\,\frac{1}{\sqrt{-
  \tilde{g}}}\,\frac{\partial }{\partial x^{\mu}}(\sqrt{-
  \tilde{g}}\,\tilde{e}^{\mu}_{\hat{\lambda}}(x))\,\gamma^{\hat{\lambda}}
+ \frac{1}{8}\,\tilde{e}^{\mu}_{\hat{\lambda}}(x)\,\tilde{\omega}_{\mu
  \hat{\alpha}\hat{\beta}}(x) [\sigma^{\hat{\alpha}\hat{\beta}},
  \gamma^{\hat{\lambda}}] - m\Big)\,\psi(x) = 0.
\end{eqnarray}
The second term in the brackets can be transformed as follows
\begin{eqnarray}\label{eq:A.9}
&&\frac{1}{2}\,i\,\frac{1}{\sqrt{- \tilde{g}}}\,\frac{\partial
  }{\partial x^{\mu}}(\sqrt{-
    \tilde{g}}\,\tilde{e}^{\mu}_{\hat{\lambda}}(x))
  \gamma^{\hat{\lambda}} = \frac{1}{2}\,i\,\frac{1}{\sqrt{-
      \tilde{g}}}\,\frac{\partial }{\partial x^{\mu}}(\sqrt{-
    \tilde{g}}\,)\,\tilde{e}^{\mu}_{\hat{\lambda}}(x)
  \gamma_{\hat{\lambda}} + \frac{1}{2}\,i\, \frac{\partial }{\partial
    x^{\mu}}\tilde{e}^{\mu}_{\hat{\lambda}}(x)\gamma^{\hat{\lambda}} =
  \nonumber\\ &&= \frac{1}{2}\,i\,\frac{1}{\sqrt{-
      \tilde{g}}}\,\frac{\partial }{\partial x^{\mu}}(\sqrt{-
    \tilde{g}}\,)\,\tilde{e}^{\mu}_{\hat{\lambda}}(x)
  \gamma_{\hat{\lambda}} + \frac{1}{2}\,i\,\Big( -
        {\tilde{\Gamma}^{\alpha}\,}_{\mu\alpha}(x)\,
        \tilde{e}^{\mu}_{\hat{\lambda}}(x)\,\gamma^{\hat{\lambda}} -
        \tilde{\omega}_{\mu\hat{\alpha}\hat{\beta}}(x)\tilde{e}^{\mu}_{\hat{\lambda}}(x)
        \eta^{\hat{\lambda}\hat{\beta}}\,\gamma^{\hat{\alpha}}\Big) = \nonumber\\ && = -
        \frac{1}{2}\,i\,{{\cal \tilde{T}}^{\alpha}\,}_{\alpha\mu}(x)
        \tilde{e}^{\mu}_{\hat{\lambda}}(x)\gamma^{\hat{\lambda}} -
        \frac{1}{2}\,i\,
        \tilde{\omega}_{\mu\hat{\alpha}\hat{\beta}}(x)\tilde{e}^{\mu}_{\hat{\lambda}}(x)
        \eta^{\hat{\lambda}\hat{\beta}}\gamma^{\hat{\alpha}},
\end{eqnarray}
where we have used the relation
\begin{eqnarray}\label{eq:A.10}
{\tilde{\Gamma}^{\alpha}\,}_{\mu\alpha}(x) =
\widetilde{\{{^{\alpha}\,}_{\mu\alpha}\}} + {\tilde{\cal
    K}^{\alpha}\,}_{\mu\alpha}(x) = \frac{1}{\sqrt{-
    \tilde{g}}}\,\frac{\partial }{\partial x^{\mu}}(\sqrt{-
  \tilde{g}}\,) + {\tilde{T}^{\alpha}\,}_{\alpha\mu}(x)
\end{eqnarray}
and the properties of the contorsion tensor ${\cal
  \tilde{K}}_{\alpha\mu\nu} = - \frac{1}{2}({\cal
  \tilde{T}}_{\alpha\mu\nu} - {\cal \tilde{T}}_{\mu\alpha\nu} - {\cal
  \tilde{T}}_{\nu\alpha\mu})$ 
\begin{eqnarray}\label{eq:A.11}
i)\;{\cal \tilde{K}}_{\alpha\nu\mu} - {\cal \tilde{K}}_{\alpha\mu\nu}
&=& {\cal
  \tilde{T}}_{\alpha\mu\nu},\nonumber\\ ii)\;\hspace{0.27in}\tilde{g}^{\mu\nu}{\cal
  \tilde{K}}_{\alpha\mu\nu} &=& - {{\cal
    \tilde{T}}^{\mu}\,}_{\mu\alpha},\nonumber\\ iii)\;\hspace{0.27in}\tilde{g}^{\alpha\mu}{\cal
  \tilde{K}}_{\alpha\mu\nu} &=&
0,\nonumber\\ iv)\;\hspace{0.27in}\tilde{g}^{\alpha\nu}{\cal
  \tilde{K}}_{\alpha\mu\nu} &=& {{\cal
    \tilde{T}}^{\alpha}\,}_{\alpha\mu}.
\end{eqnarray}
Thus, the Dirac equation in the curved spacetime takes the form
\begin{eqnarray}\label{eq:A.12}
\Big( i\,\tilde{e}^{\mu}_{\hat{\lambda}}(x) \gamma^{\hat{\lambda}}
D_{\mu} - \frac{1}{2}\,i\,{{\cal
    \tilde{T}}^{\alpha}\,\!\!}_{\alpha\mu}(x)
\tilde{e}^{\mu}_{\hat{\lambda}} (x) \gamma^{\hat{\lambda}} -
\frac{1}{2}\,i\,
\tilde{\omega}_{\mu\hat{\alpha}\hat{\beta}}(x)\tilde{e}^{\mu}_{\hat{\lambda}}(x)
\Big(\eta^{\hat{\lambda}\hat{\beta}}\gamma^{\hat{\alpha}} +
\frac{1}{4}\,i\, [\sigma^{\hat{\alpha}\hat{\beta}},
  \gamma^{\hat{\lambda}}]\Big) - m\Big)\,\psi(x) = 0
\end{eqnarray}
and agrees well with Eq.(\ref{eq:18}) of
Ref.\cite{Kostelecky2004}. From Eq.(\ref{eq:A.12}) we derive the Dirac
Hamilton operator
\begin{eqnarray}\label{eq:A.13}
{\rm H} = {\rm H}_0 + \delta {\rm H},
\end{eqnarray}
where ${\rm H}_0 = \gamma^{\hat{0}} m -
i\,\gamma^{\hat{0}}\,\hat{\!\vec{\gamma}}\cdot \vec{\nabla}$ and $\delta
{\rm H}$ is given by
\begin{eqnarray}\label{eq:A.14}
\delta {\rm H} &=& (\tilde{e}^{\hat{0}}_0(x) - 1)\gamma^{\hat{0}} m -
i\,(\tilde{e}^{\hat{0}}_0(x) -
1)\gamma^{\hat{0}}\gamma^{\hat{j}}\delta^j_{\hat{j}}\frac{\partial
}{\partial x^j} - i \tilde{e}^{\hat{0}}_0(x)(\tilde{e}^j_{\hat{j}}(x) -
\delta^j_{\hat{j}})\gamma^{\hat{0}}\gamma^{\hat{j}}\frac{\partial
}{\partial x^j}\nonumber\\ &+& \frac{1}{2}\,i\,{\tilde{\cal
    T}^{\alpha}\,\!\!}_{\alpha\mu}(x) \tilde{e}^{\hat{0}}_0(x)
\tilde{e}^{\mu}_{\hat{\lambda}}(x)
\gamma^{\hat{0}}\gamma^{\hat{\lambda}} + \frac{1}{2}\,i\,
\tilde{\omega}_{\mu\hat{\alpha}\hat{\beta}}(x) \tilde{e}^{\hat{0}}_0(x)
\tilde{e}^{\mu}_{\hat{\lambda}}(x)\gamma^{\hat{0}}
\Big(\eta^{\hat{\lambda}\hat{\beta}}\gamma^{\hat{\alpha}} +
\frac{1}{4}\,i\,
\{\sigma^{\hat{\alpha}\hat{\beta}},\gamma^{\hat{\lambda}}\}\Big),
\end{eqnarray}
which is valid for Dirac fermions in curved spacetimes
  with diagonal metric tensors, related to the vierbein fields with
  vanishing non--diagonal time--space (space--time) components.  For
the calculation of the Hamilton operator Eq.(\ref{eq:A.14}) it is
convenient to use the following relations
\begin{eqnarray}\label{eq:A.15}
\gamma^{\hat{\lambda}}\sigma^{\hat{\alpha}\hat{\beta}} &=&
i\,(\eta^{\hat{\lambda}\hat{\alpha}}\,\gamma^{\hat{\beta}} -
\eta^{\hat{\beta}\hat{\lambda}}\,\gamma^{\hat{\alpha}}) -
\epsilon^{\hat{\lambda}\hat{\alpha}\hat{\beta}\hat{\rho}}\gamma_{\hat{\rho}}
\gamma^5,\nonumber\\ \sigma^{\hat{\alpha}\hat{\beta}}\gamma^{\hat{\lambda}}
&=& i\,(\eta^{\hat{\beta}\hat{\lambda}}\,\gamma^{\hat{\alpha}} -
\eta^{\hat{\lambda}\hat{\alpha}}\,\gamma^{\hat{\beta}}) -
\epsilon^{\hat{\lambda}\hat{\alpha}\hat{\beta}\hat{\rho}}\gamma_{\hat{\rho}}
\gamma^5,\nonumber\\ \{\sigma^{\hat{\alpha}\hat{\beta}},
\gamma^{\hat{\lambda}}\} &=& - 2\,
\epsilon^{\hat{\lambda}\hat{\alpha}\hat{\beta}\hat{\rho}}\gamma_{\hat{\rho}}
\gamma^5.
\end{eqnarray}
 Substituting Eq.(\ref{eq:A.15}) into Eq.(\ref{eq:A.14}) we arrive at
 the following Hamilton operator
\begin{eqnarray}\label{eq:A.16}
\delta {\rm H} &=& (\tilde{e}^{\hat{0}}_0(x) - 1)\gamma^{\hat{0}} m -
i\,(\tilde{e}^{\hat{0}}_0(x) -
1)\gamma^{\hat{0}}\gamma^{\hat{j}}\delta^j_{\hat{j}}\frac{\partial
}{\partial x^j} - i\tilde{e}^{\hat{0}}_0(x)(\tilde{e}^j_{\hat{j}}(x) -
\delta^j_{\hat{j}})\gamma^{\hat{0}}\gamma^{\hat{j}}\frac{\partial
}{\partial x^j}\nonumber\\&+& \frac{1}{2}\,i\,{\tilde{\cal
    T}^{\alpha}\,\!\!}_{\alpha\mu}(x)\tilde{e}^{\hat{0}}_0(x)
\tilde{e}^{\mu}_{\hat{\lambda}}(x)
\gamma^{\hat{0}}\gamma^{\hat{\lambda}} + \frac{1}{2}\,i\,
\tilde{\omega}_{\mu\hat{\alpha}\hat{\beta}}(x)\tilde{e}^{\hat{0}}_0(x)
\tilde{e}^{\mu}_{\hat{\lambda}}(x)\gamma^{\hat{0}}
\Big(\eta^{\hat{\lambda}\hat{\beta}}\gamma^{\hat{\alpha}} -
\frac{1}{2}\,i\,\epsilon^{\hat{\lambda}\hat{\alpha}\hat{\beta}\hat{\rho}}
\gamma_{\hat{\rho}} \gamma^5 \Big).
\end{eqnarray}
Then, the spin connection $\tilde{\omega}_{\mu\hat{\alpha}\hat{\beta}}(x)$ is defined in terms of the vierbein fields and the affine connection as follows \cite{Kostelecky2004}
\begin{eqnarray}\label{eq:A.17}
\tilde{\omega}_{\mu\hat{\alpha}\hat{\beta}}(x) = -
\eta_{\hat{\alpha}\hat{\varphi}}\Big(\partial_{\mu}\tilde{e}^{\hat{\varphi}}_{\nu}(x)
- {\Gamma^{\alpha}}_{\mu\nu}(x)
\tilde{e}^{\hat{\varphi}}_{\alpha}(x)\Big)\tilde{e}^{\nu}_{\hat{\beta}}(x).
\end{eqnarray}
As a result, the Hamilton operator Eq.(\ref{eq:A.16}) reads
\begin{eqnarray}\label{eq:A.18}
\hspace{-0.3in}\delta {\rm H} &=& (\tilde{e}^{\hat{0}}_0(x) - 1)\gamma^{\hat{0}} m -
i\,(\tilde{e}^{\hat{0}}_0(x) -
1)\gamma^{\hat{0}}\gamma^{\hat{j}}\delta^j_{\hat{j}}\frac{\partial
}{\partial x^j} - i\tilde{e}^{\hat{0}}_0(x)(\tilde{e}^j_{\hat{j}}(x) -
\delta^j_{\hat{j}})\gamma^{\hat{0}}\gamma^{\hat{j}}\frac{\partial
}{\partial x^j}\nonumber\\
\hspace{-0.3in}&+& \frac{1}{2}\,i\,{\tilde{\cal
    T}^{\alpha}\,\!\!}_{\alpha\mu}(x) \tilde{e}^{\hat{0}}_0(x)
\tilde{e}^{\mu}_{\hat{\lambda}}(x)
\gamma^{\hat{0}}\gamma^{\hat{\lambda}} - \frac{1}{2}\,i\,
\eta_{\hat{\alpha}\hat{\varphi}}\Big(\partial_{\mu}\tilde{e}^{\hat{\varphi}}_{\nu}(x)
- {\Gamma^{\alpha}}_{\mu\nu}(x)
\tilde{e}^{\hat{\varphi}}_{\alpha}(x)\Big)\tilde{e}^{\nu}_{\hat{\beta}}(x)
\tilde{e}^{\hat{0}}_0(x) \tilde{e}^{\mu}_{\hat{\lambda}}(x)\nonumber\\
\hspace{-0.3in}&\times&\gamma^{\hat{0}}
\Big(\eta^{\hat{\lambda}\hat{\beta}}\gamma^{\hat{\alpha}} -
\frac{1}{2}\,i\,\epsilon^{\hat{\lambda}\hat{\alpha}\hat{\beta}\hat{\rho}}
\gamma_{\hat{\rho}} \gamma^5 \Big).
\end{eqnarray}
Keeping only the linear contributions of interacting fields we rewrite
the Hamilton operator $\delta {\rm H}$ as follows
\begin{eqnarray}\label{eq:A.19}
\hspace{-0.3in}\delta {\rm H} &=& (\tilde{e}^{\hat{0}}_0(x) - 1)
\gamma^{\hat{0}} m - i\,(\tilde{e}^{\hat{0}}_0(x) -
1)\gamma^{\hat{0}}\gamma^{\hat{j}}\delta^j_{\hat{j}}\frac{\partial
}{\partial x^j} - i\tilde{e}^{\hat{0}}_0(x)(\tilde{e}^j_{\hat{j}}(x) -
\delta^j_{\hat{j}})\gamma^{\hat{0}}\gamma^{\hat{j}}\frac{\partial
}{\partial x^j} -
\frac{1}{2}\,i\,\eta^{\mu\nu}\Big(\partial_{\mu}\tilde{e}^{\hat{\lambda}}_{\nu}(x)
-
\{{^{\hat{\lambda}}}_{\mu\nu}\}\Big)\gamma^{\hat{0}}\gamma_{\hat{\lambda}}\nonumber\\
\hspace{-0.3in}&+& \frac{1}{2}\,i\,{\tilde{\cal
    T}^{\alpha}\,\!\!}_{\alpha\hat{\lambda}}(x)
\gamma^{\hat{0}}\gamma^{\hat{\lambda}} +
\frac{1}{2}\,i\,\eta^{\mu\nu} {\cal K}_{\hat{\lambda}\mu\nu}(x)
\gamma^{\hat{0}}\gamma^{\hat{\lambda}} +
\frac{1}{4}\,\epsilon^{\hat{\alpha}\hat{\beta}\mu\nu}{\cal
  K}_{\hat{\beta}\mu\nu}\gamma^{\hat{0}}\gamma_{\hat{\alpha}}\gamma^5.
\end{eqnarray}
Using the properties of the contorsion tensor Eq.(\ref{eq:A.11}) we
get
\begin{eqnarray}\label{eq:A.20}
\hspace{-0.3in}\delta {\rm H} &=& (\tilde{e}^{\hat{0}}_0(x) - 1)
\gamma^{\hat{0}} m - i\,(\tilde{e}^{\hat{0}}_0(x) -
1)\gamma^{\hat{0}}\gamma^{\hat{j}}\delta^j_{\hat{j}}\frac{\partial
}{\partial x^j} - i\tilde{e}^{\hat{0}}_0(x)(\tilde{e}^j_{\hat{j}}(x) -
\delta^j_{\hat{j}})\gamma^{\hat{0}}\gamma^{\hat{j}}\frac{\partial
}{\partial x^j} -
\frac{1}{2}\,i\,\eta^{\mu\nu}\Big(\partial_{\mu}\tilde{e}^{\hat{\lambda}}_{\nu}(x)
-
\{{^{\hat{\lambda}}}_{\mu\nu}\}\Big)\gamma^{\hat{0}}\gamma_{\hat{\lambda}}\nonumber\\
\hspace{-0.3in}&-&
\frac{1}{8}\,\epsilon^{\hat{\alpha}\hat{\beta}\mu\nu}{\cal
  T}_{\hat{\beta}\mu\nu}\gamma^{\hat{0}}\gamma_{\hat{\alpha}}\gamma^5.
\end{eqnarray}
The term $\frac{1}{2}\,i\,{\tilde{\cal
    T}^{\alpha}\,\!\!}_{\alpha\hat{\lambda}}(x)
\gamma^{\hat{0}}\gamma^{\hat{\lambda}}$ is cancelled in agreement with
the analysis of the Dirac equation, carried out by Kostelecky
\cite{Kostelecky2004}. This implies that the axial 4--vector torsion
components can possess only the minimal torsion--fermion couplings for
the Dirac fermions.

\end{document}